\documentclass[reprint,aps,prb,a4,superscriptaddress]{revtex4-1}

\usepackage{graphicx}
\usepackage{dcolumn,amsmath,amssymb,amsfonts}
\usepackage{bm}
\usepackage{color}
\usepackage[dvipsnames]{xcolor}

\begin{document}
\title{The electrocaloric effect in BaTiO$_3$ at all three
  ferroelectric transitions: anisotropy and inverse caloric effects}

\author{Madhura Marathe} \email{madhura.marathe@mat.ethz.ch}
\affiliation{Materials Theory, ETH Z\"urich, Wolfgang-Pauli-Str. 27,
  8093 Z\"urich, Switzerland}

\author{Damian Renggli} 
\affiliation{Materials Theory, ETH Z\"urich, Wolfgang-Pauli-Str. 27,
  8093 Z\"urich, Switzerland}

\author{Mehmet Sanlialp}
\affiliation{Institute for Materials Science and Center for 
  Nanointegration Duisburg-Essen (CENIDE), University of 
  Duisburg-Essen, 45141 Essen, Germany}

\author{Maksim O. Karabasov}
\affiliation{Institute for Materials Science and Center for 
  Nanointegration Duisburg-Essen (CENIDE), University of 
  Duisburg-Essen, 45141 Essen, Germany}

\author{Vladimir V. Shvartsman}
\affiliation{Institute for Materials Science and Center for 
  Nanointegration Duisburg-Essen (CENIDE), University of 
  Duisburg-Essen, 45141 Essen, Germany}

\author{Doru C. Lupascu}
\affiliation{Institute for Materials Science and Center for 
  Nanointegration Duisburg-Essen (CENIDE), University of 
  Duisburg-Essen, 45141 Essen, Germany}

\author{Anna Gr\"unebohm} 
\affiliation{Faculty of Physics and Center for Nanointegration 
  Duisburg-Essen (CENIDE), University of Duisburg-Essen, 47048, 
  Duisburg, Germany}

\author{Claude Ederer} \email{claude.ederer@mat.ethz.ch}
\affiliation{Materials Theory, ETH Z\"urich, Wolfgang-Pauli-Str. 27,
  8093 Z\"urich, Switzerland}

\date{\today}

\begin{abstract}
We study the electrocaloric (EC) effect in bulk BaTiO$_3$ (BTO) using
molecular dynamics simulations of a first principles-based effective
Hamiltonian, combined with direct measurements of the adiabatic EC
temperature change in BTO single crystals. We examine in particular the
dependence of the EC effect on the direction of the applied electric
field at all three ferroelectric transitions, and we show that the EC
response is strongly anisotropic. Most strikingly, an inverse caloric
effect, i.e., a temperature increase under field removal, can be
observed at both ferroelectric-ferroelectric transitions for certain
orientations of the applied field. Using the generalized
Clausius-Clapeyron equation, we show that the inverse effect occurs
exactly for those cases where the field orientation favors the higher
temperature/higher entropy phase. Our simulations show that
temperature changes of around 1\,K can in principle be obtained at the
tetragonal-orthorhombic transition close to room temperature, even for
small applied fields, provided that the applied field is strong enough
to drive the system across the first order transition line.
Our direct EC measurements for BTO single crystals at the
cubic-tetragonal and at the tetragonal-orthorhombic transitions are in
good qualitative agreement with our theoretical predictions, and in
particular confirm the occurrence of an inverse EC effect at the
tetragonal-orthorhombic transition for electric fields applied along
the [001] pseudo-cubic direction.
\end{abstract}

\maketitle

\section{Introduction}

The electrocaloric (EC) effect -- a reversible temperature change of a
material under adiabatic application or removal of an electric field
-- was first reported in 1930.~\cite{Kobeco_Kurtchatov:1930} While
initially the effect was considered too weak to be useful for
applications, technological applications are now considered feasible,
after a very large EC temperature change of $\sim$12\,K has been
reported for thin films of
PbZr$_{0.95}$Ti$_{0.05}$O$_3$.~\cite{Mischenko_et_al_2006} This
observation has led to a large increase in research activity on the EC
effect.~\cite{Scott_2011,Valant_2012,Liu/Scott/Dkhil:2016} The EC
effect, as well as the analogous magnetocaloric and elastocaloric
effects, which can be observed in certain materials under application
of a magnetic field or a stress field,
respectively,~\cite{Manosa/Planes/Acet:2013,Moya/Narayan/Mathur:2014}
are currently attracting considerable attention. All three caloric
effects can facilitate the development of a new generation of solid
state cooling devices, which promise to be more energy-efficient and
environmentally-friendly than currently existing devices based on
vapor compression.~\cite{Faehler_et_al:2011,Takeuchi/Sandeman:2015}

In the present work, we focus on BaTiO$_3$ (BTO), which is a 
well-characterized prototypical ferroelectric (FE) material. 
BTO exhibits a paraelectric (PE) cubic (C) phase at high temperature,
which on cooling transforms into a FE tetragonal (T) phase at $\sim
120^\circ$C. Further cooling leads to two FE-FE transitions, first to
an orthorhombic (O) phase at $\sim 5^\circ$C and finally to a
rhombohedral (R) phase at $\sim -90^\circ$C.~\cite{Lines-Glass} In
each FE phase, the spontaneous electric polarization points along a
different crystallographic direction. The strongest EC effect is 
typically observed in FE materials close to a FE phase transition, 
where application of a moderate electric field can result in very 
large changes of the electric polarization.~\cite{Valant_2012} 
Therefore, the various transitions make BTO a very attractive system 
for exploring the EC effect. 

The EC effect in BTO has been studied both experimentally, see e.g.,
Refs.~\onlinecite{Narayan/Mathur:2010,Moya_et_al_2013,Novak_Pirc_Kutnjak:2013},
as well as theoretically, using either phenomenological
thermodynamical modeling or an \textit{ab initio}-based effective
Hamiltonian approach, see e.g.,
Refs.~\onlinecite{Akcay_et_al_2007,Akcay_et_al_2008,Novak_Kutnjak_Pirc:2013,Beckman_et_al_2012,Nishimatsu_Barr_Beckman:2013,Marathe/Ederer:2014,Grunebohm_Nishimatsu:2016}.
Most of these previous studies have focused on the temperature region
around the PE-FE transition, where particularly large field-induced
polarization changes occur. However, large polarization changes,
including reorientation of the polarization direction along different
crystallographic axes, also occur at the FE-FE transitions. Indeed, an
EC effect corresponding to a temperature change of 1.4\,K under
application of an electric field of 10\,kV/cm has been reported at the
T-O transition in BTO.~\cite{Bai_et_al:2012} Furthermore, \textit{ab
  initio}-based studies predict sizable EC temperature changes both
at the T-O and at the O-R transition in
BTO,~\cite{Nishimatsu_Barr_Beckman:2013} as well as in the closely
related compound
Ba$_{0.5}$Sr$_{0.5}$TiO$_3$.~\cite{Ponomareva/Lisenkov:2012} A finite
EC response has also been measured at FE-FE transitions in other
ferroelectrics, e.g., Pb(Mg$_{1/3}$Nb$_{2/3}$)O$_3$-PbTiO$_3$ single
crystals.~\cite{Perantie_et_al:2010}

Generally, the EC effect is related to the polarization change along
the direction of the applied field, and thus most theoretical studies
have considered an electric field applied along the direction of the
spontaneous polarization in the FE phase. Nevertheless, some
experimental studies have been performed for Pb-based relaxor single
crystals grown with different
orientations,~\cite{Sebald_et_al_2006,Luo_et_al_2012,Chukka_et_al_2013}
and have confirmed that the EC response indeed depends on the
orientation of the applied field. This \emph{anisotropy} of the EC
effect has not received much attention so far. In the case of a FE-FE
transition for which the orientation of the spontaneous polarization
changes between two crystallographically inequivalent directions, the
anisotropy of the EC effect is particularly relevant, and is currently
not fully understood.

Remarkably, an \emph{inverse} (or negative) EC response, i.e., a
temperature decrease under application of an electric field, has been
observed for certain applied field directions within a small
temperature region.~\cite{Perantie_et_al:2010} Ponomareva and Lisenkov
have attributed this inverse EC effect to noncollinearity between the
electric polarization and the applied field, and were able to
reproduce the inverse effect in their \textit{ab initio}-based
simulations.~\cite{Ponomareva/Lisenkov:2012} However, a full
understanding of this inverse EC effect and the conditions required
for observing it at FE-FE transitions is still lacking.

Due to its multiple FE transitions, BTO is an ideal system to develop
a better general understanding of the EC response at FE-FE
transitions, of the corresponding anisotropy, and of a possible
inverse EC response. We note that the first FE-FE (T-O) transition in
BTO occurs just below room temperature, and thus in a very attractive
temperature region for many anticipated technological
applications. Furthermore, modified EC cycles have been suggested
which could enhance the overall EC response of a material by utilizing
a combination of normal and inverse EC
effects.~\cite{Ponomareva/Lisenkov:2012,Yang-bin_Ma_et_al:2016}

In this paper, we present a detailed and systematic study of the EC
effect in BTO at all three FE transitions using a
first principles-based effective Hamiltonian approach.  In particular,
we examine how the direction of the applied field affects the EC
temperature change. Since application of an external electric field
shifts the phase transition temperatures and can also affect the order
of the corresponding phase transition, we first establish the
electric-field versus temperature phase diagram for different
orientations of the electric field. We obtain a finite adiabatic EC
temperature change at all three transitions. For some field
orientations an inverse EC effect can be observed at the FE-FE
transitions, and we analyze the mechanism leading to such inverse
effects in terms of the Clausius-Clapeyron equation. To verify our
theoretical predictions and the results of our simulations, we also
perform direct measurements of the EC temperature change in BTO single
crystals along different crystallographic directions. We observe good
qualitative agreement between our simulations and the experimental
measurements.

This paper is organized as follows. In Secs.~\ref{sec:method}
and~\ref{sec:exp}, we describe our computational and experimental
methods, respectively. We then present and discuss the calculated
electric field versus temperature phase diagrams
(Sec.~\ref{subsec:ET-pd}), followed by the discussion of potential
inverse caloric effects using the generalized Clausius-Clapeyron
equation, along with an estimation of the entropy changes associated
with the latent heat of the various first-order phase transitions
(Sec.~\ref{subsec:cc}). Next, we present the calculated EC temperature
changes for different applied field directions, and we also discuss
some examples of non-monotonic behavior of the EC temperature changes
as function of the electric field magnitude
(Sec.~\ref{subsec:deltaT}). Finally, in Sec.~\ref{subsec:exptdeltaT}
we present the directly measured EC temperature changes for BTO single
crystals corresponding to different orientations of the applied field,
and we summarize our results in Sec.~\ref{sec:summary}.

\section{Computational Method}
\label{sec:method}

We use a first principles-based effective Hamiltonian
approach,~\cite{Zhong_Vanderbilt_Rabe_1994,Zhong_Vanderbilt_Rabe_1995}
which is applicable to FE materials exhibiting a cubic perovskite
structure at high temperatures. Within this approach, only those
degrees of freedom are considered that are essential to correctly
describe the FE transitions. To this end, each cubic perovskite unit
cell, $i$, is represented by a three-dimensional soft mode vector
$\mathbf{u}_i$ and a vector describing the local strain
$\mathbf{w}_i$.  Further, the global elastic degrees of freedom are
included through the homogeneous strain tensor $\eta_j$, $j=1, \dots,
6$ (in standard Voigt notation).  The total energy of the system is
then expressed as a low order polynomial in terms of these variables,
and includes various terms describing the so-called soft-mode self
energy (the tendency to form local dipoles), the dipole-dipole
interaction energy, a short range interaction between soft modes, the
homogeneous and inhomogeneous elastic energy, and additional terms
describing the coupling between soft mode and strain variables. The
latter are essential to correctly describe the three successive phase
transitions in BTO and related materials.  All parameters of this
effective Hamiltonian can be calculated using \textit{ab initio}
density functional theory,
~\cite{Zhong_Vanderbilt_Rabe_1995,Nishimatsu_et_al_2010} i.e., no
empirical fitting to experimental data is required.

We use the open-source ``feram'' code
(\url{http://loto.sourceforge.net/feram/})~\cite{Nishimatsu_et_al_2008}
to perform molecular dynamics (MD) simulations for this effective
Hamiltonian, allowing us to access finite temperature
properties. Further, we use the previously reported parameter set for
bulk BaTiO$_3$,~\cite{Nishimatsu_et_al_2010} determined using density
functional theory calculations with the Wu-Cohen exchange-correlation
functional.~\cite{Wu/Cohen:2006} A $96 \times 96 \times 96$ supercell
is used in all our simulations, corresponding to about 900\,000
perovskite units. The system is equilibrated at a given temperature,
with or without applied field, within the canonical, i.e., ``constant
temperature'' (NPT) ensemble using a Nos\'e-Poincar\'e
thermostat.~\cite{Bond/Leimkuhler/Laird:1999}

The transition temperatures for the FE phase transitions at a fixed
applied electric field are determined sequentially by performing both
``heating'' and ``cooling'' simulations, i.e., each simulation is
initialized using a previously thermalized configuration obtained at
slightly lower or higher temperature, respectively. We use temperature
steps of 10\,K far away from a phase transition and reduced steps of
2\,K near phase transitions. The system is thermalized over a time
period of 120\,ps, and then statistical averages for various
quantities are accumulated over a period of 160\,ps, using MD time
steps of 2\,fs.

\begin{figure}[tb]
\centering
\includegraphics[angle=270,width=0.4\textwidth]{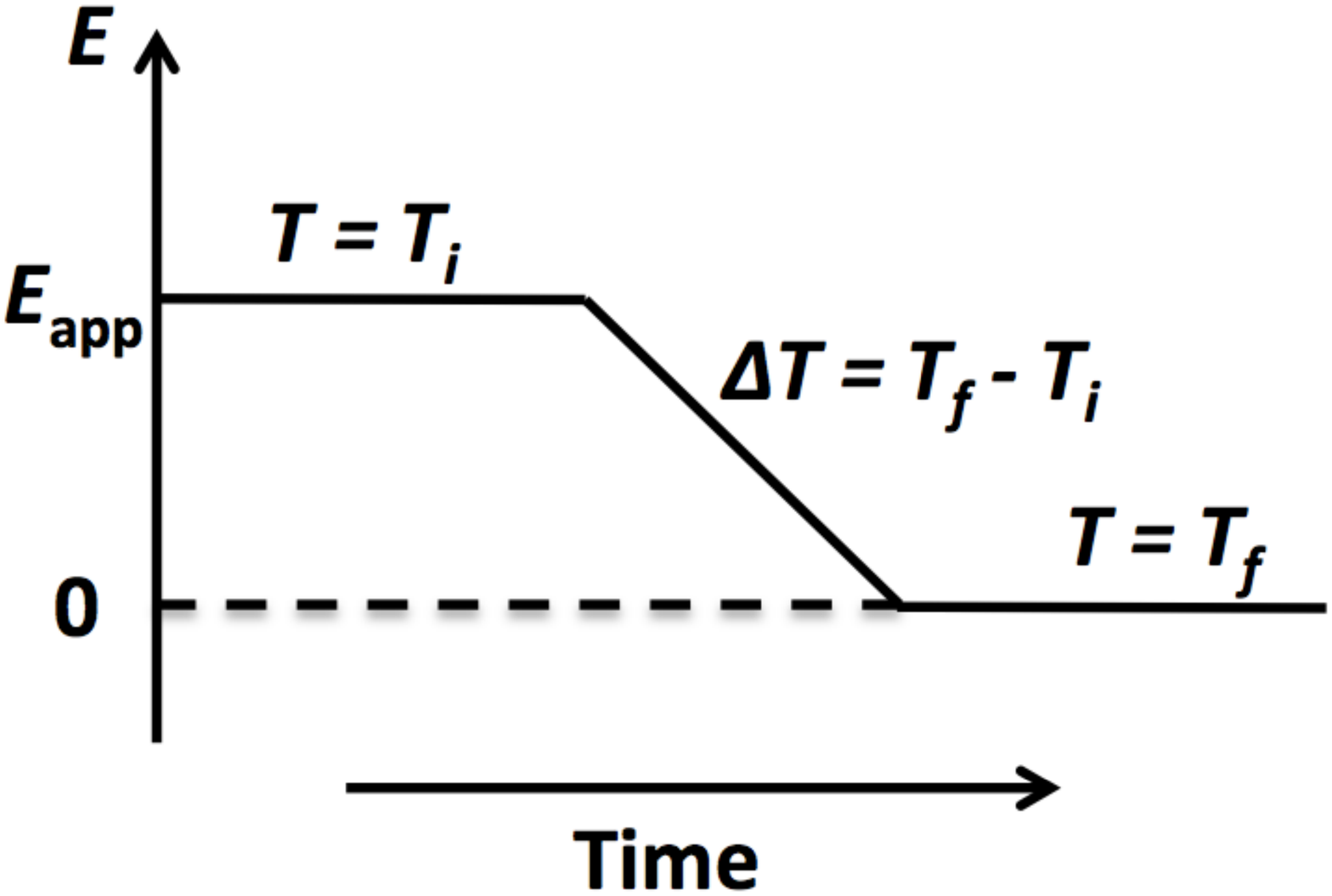}
\caption{Schematic representation of the computational protocol used
  to obtain the adiabatic temperature change for switching the field
  off. Here, $E_{\text{app}}$ is the applied field. $T_i$ and $T_f$
  denote initial and final temperatures corresponding to
  $E_{\text{kin}}^{i}$ and $E_{\text{kin}}^{f}$, respectively.}
\label{fig:delT-scheme}
\end{figure}

To determine the adiabatic temperature change under application or
removal of an external electric field, we employ the microcanonical,
i.e., ``constant energy'' (NPE) ensemble after the system has been
thermalized at a given temperature and field.  This allows us to
monitor the change in kinetic energy of the system while the electric
field is slowly ramped up or down. The EC temperature change $\Delta
T$ is then obtained using:
\begin{equation}
\Delta T = \frac{2(E_{\text{kin}}^{f} - E_{\text{kin}}^{i})}{N_fk_B}. 
\end{equation}
Here, $E_{\text{kin}}^{i}$ and $E_{\text{kin}}^{f}$ denote the initial
and final average kinetic energies of the system, $N_f$ is the number
of degrees of freedom of the model system, and $k_B$ is the Boltzmann
constant. The computational protocol is schematically shown in
Fig.~\ref{fig:delT-scheme}. Except where otherwise noted, the applied
electric field along each Cartesian direction is changed with a rate
0.002\,kV/(cm$\cdot$fs), which is slow enough to ensure that the
system remains in thermal equilibrium during the field
ramping.~\cite{Marathe_et_al:2016} The system is thermalized before
and after the ramping for 80\,ps, and then thermal averages are
accumulated over a period of 40\,ps (before the field ramping) and
100\,ps (after the field ramping).  A MD time step of 1\,fs is used
for these simulations.

We note that, for computational efficiency, only the soft mode
variables are treated as dynamical variables within this work, whereas
the local and global strain variables are obtained by minimizing the
total energy for the current soft mode configuration in each MD
step. Thus, our simulations contain only 3 dynamical (fluctuating)
variables per perovskite unit cell, as compared to 15 in the real
material. As a result, the specific heat of the model system is
smaller than that of the real material by approximately a factor of
$3/15=1/5$. Consequently, the calculated $\Delta T$ is overestimated
by approximately a factor of $5$, and in the following we always
report ``scaled'' $\Delta T$ values, which are corrected by this
factor to match our values to the number of degrees of freedom of the
real system.~\cite{Nishimatsu_Barr_Beckman:2013} In contrast, whenever
we report absolute temperature values, these correspond to the actual
system temperature $T$ of our model system during the simulation.  In
this case, rescaling is not applicable.

Further, we note that, while the effective Hamiltonian approach is
able to successfully reproduce all three FE transitions in
BTO,\cite{Zhong_Vanderbilt_Rabe_1995,Nishimatsu_et_al_2010} the
calculated transition temperatures deviate by some amount from the
experimentally measured values. These deviations result from the
simplifications inherent in the effective Hamiltonian method (reduced
number of variables, neglect of higher order terms in the total
energy) as well as from the approximations used to calculate the
corresponding parameters using density functional
theory.\cite{Tinte_et_al:2003} While some of these deviations can be
reduced by using either fixed or temperature-dependent pressure
corrections (see, e.g.,
Refs.~\onlinecite{Zhong_Vanderbilt_Rabe_1995,Nishimatsu_et_al_2010}),
in the present work we do not use such empirical corrections, since
their applicability is not apparent if the temperature of the system
varies during the simulation.~\cite{Nishimatsu_Barr_Beckman:2013} We
point out that the effective Hamiltonian approach was shown to be a
powerful method that allows to obtain general understanding and give
crucial information on overall magnitudes as well as expected trends.

\section{Experimental Method}
\label{sec:exp}

BTO single crystals (3\,mm $\times$ 3\,mm $\times$ 0.5\,mm) cut
perpendicular to the [001], [011] and [111] directions, were purchased
from EQ Photonics GmbH.  For electrocaloric measurements they were
sputtered with 100 nm Au electrodes on both faces. We note that here
and in the following, we always use cubic (or pseudo-cubic) notation
to specify the crystallographic directions, i.e. [100], [010], and
[001] correspond to the Cartesian $x$, $y$, and $z$ directions.

The EC effect was directly measured using a custom-built
quasi-adiabatic calorimeter under high vacuum-condition (about
$10^{-6}$\,mbar). The measurements were performed on heating between
276\,K and 420\,K. Electric field pulses with magnitudes of 5, 7.5,
and 10\,kV/cm, and a period of 200\,s, were applied at each
measurement temperature. For each temperature and field value, four
temperature changes were measured over two field cycles (i.e., two
measurements for switching on and two for switching off the
field). From this data, the mean EC temperature change and the
corresponding standard deviation were obtained.  The electrical pulses
were generated by a functional signal generator (Keithley Model 3390)
and amplified by a high-voltage amplifier (TREK Model PD05034). The
sample temperature change, $\Delta T_{\text{meas}}$, was recorded by a
Kapton$^{\textregistered}$ - insulated type K thermocouple, contacted
to the top electrode of the sample and connected to a temperature
controller (Lakeshore Cryotronics Model 336). The EC temperature
change was then calculated taking into account the geometry of the
system and the heat capacities of its components:
\begin{equation}
\Delta T = \Delta T_{\text{meas}} {\sum_i  \frac{C_p^i}{C_p^{EC}}}. 
\end{equation}
Here, $C_p^i$ represent the heat capacities of each subsystem $i$
(parts of the sample with electrode and without electrode, electrical
wires, alumina compound to contact the thermocouple, silver paste to
fix the electrical wires) which are in contact with the sample, and
$C_p^{EC}$ is the heat capacity of the BTO single crystal covered by
the electrode.~\cite{Rozic_et_al_2010}
Both the heat capacity and the phase transition temperatures of the
BTO single crystals were measured using a differential scanning
calorimeter (Netzsch, DSC 204) on heating and on cooling at 10\,K/min
over a temperature range of 250-425\,K, as reported
elsewhere.~\cite{Sanlialp_et_al_2016}

\section{Results and Discussion}
\label{sec:results}

\subsection{$E$-$T$ phase diagrams}
\label{subsec:ET-pd}

\begin{figure}
\centering
\includegraphics[width=0.4\textwidth]{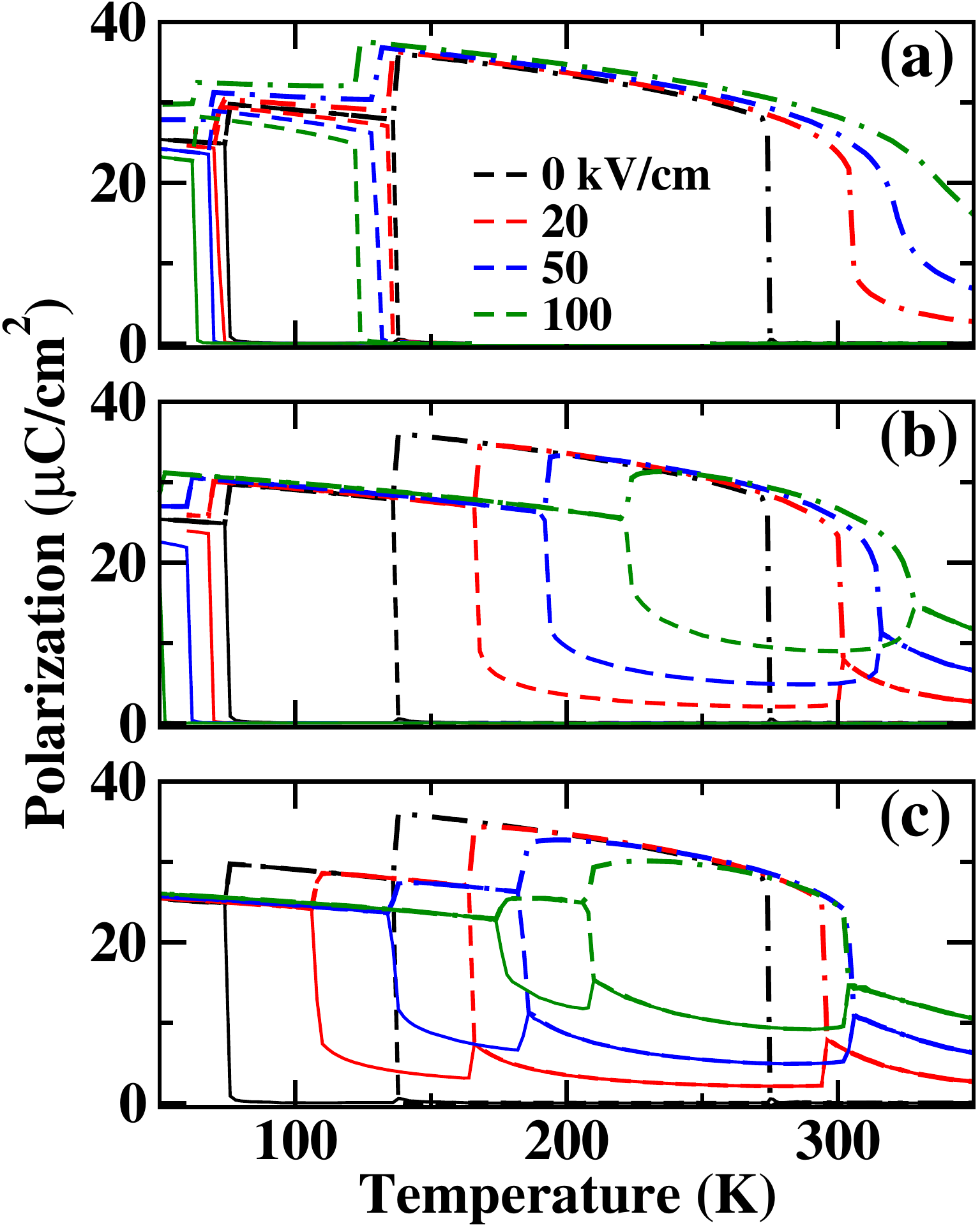}
\caption{Polarization components along $x$ (solid lines), $y$ (dashed 
  lines), and $z$ (dot-dashed lines) directions are plotted as a 
  function of temperature for several electric fields applied along 
  (a) [001], (b) [011] and (c) [111]. For clarity, only the results 
  obtained from cooling simulations and for selected field strengths 
  $E_\alpha$ are shown.}
\label{fig:P_T-Eapp}
\end{figure}

We first examine how the magnitude as well as the direction of an
applied electric field alters the nature of the different phases and
the corresponding phase transition temperatures in bulk BaTiO$_3$. We
consider three different field directions: [001], [011], and [111],
which correspond to the directions of the spontaneous polarization in
the T, O, and R phases, respectively. We determine the phase
transition temperatures from the temperature dependence of the
different polarization components calculated during heating and
cooling simulations. As example, different components of the
polarization, $P_\alpha$, $\alpha \in \{x, y, z\}$, are plotted as a
function of temperature for a few applied fields in
Fig.~\ref{fig:P_T-Eapp}. For brevity, we only show data from cooling
simulations. The data obtained from heating simulations is analogous,
except that the resulting transition temperatures are typically
somewhat higher than the ones obtained from the cooling
simulations. This is due to the thermal hysteresis associated with the
first order phase transitions.  Note that here and in the following,
we usually specify the magnitude $E_\alpha$ for each nonzero component
of the applied field, i.e., the corresponding fields $\vec{E}$ are
$(0, 0, E_\alpha)$, $(0, E_\alpha, E_\alpha)$, and $(E_\alpha,
E_\alpha, E_\alpha)$ for fields applied along [001], [011], and [111],
respectively.

For zero electric field, sharp jumps can be observed in the different
polarization components at the transition temperatures (see black
lines in Fig.~\ref{fig:P_T-Eapp}), indicating the first order
(discontinuous) character of each transition, as expected for bulk
BTO.~\cite{Lines-Glass} With decreasing temperature, one can observe
successive transitions from the PE-C phase to the FE-T, O, and R
phases, corresponding to one, two, and three nonzero Cartesian
components of the polarization, respectively.

Under application of an electric field, the transition temperatures
shift to either lower or higher temperatures, depending on the
specific transition and field direction.  Simultaneously, in all cases
where the electric field is not parallel to the zero-field
polarization direction, the symmetry of the corresponding phase is
reduced to monoclinic, due to non-zero polarization components induced
by the applied field, or due to enhanced polarization components along
the field direction.  In spite of this symmetry-lowering, in the
following we will continue to label the different FE phases as T, O,
and R, i.e., according to their zero field symmetries. Furthermore,
for each field direction the symmetry of the ``PE'' phase, becomes
identical to the one of the FE phase that has its spontaneous
zero-field polarization along the direction of the applied field.

\begin{figure}
\centering \includegraphics[width=0.4\textwidth]{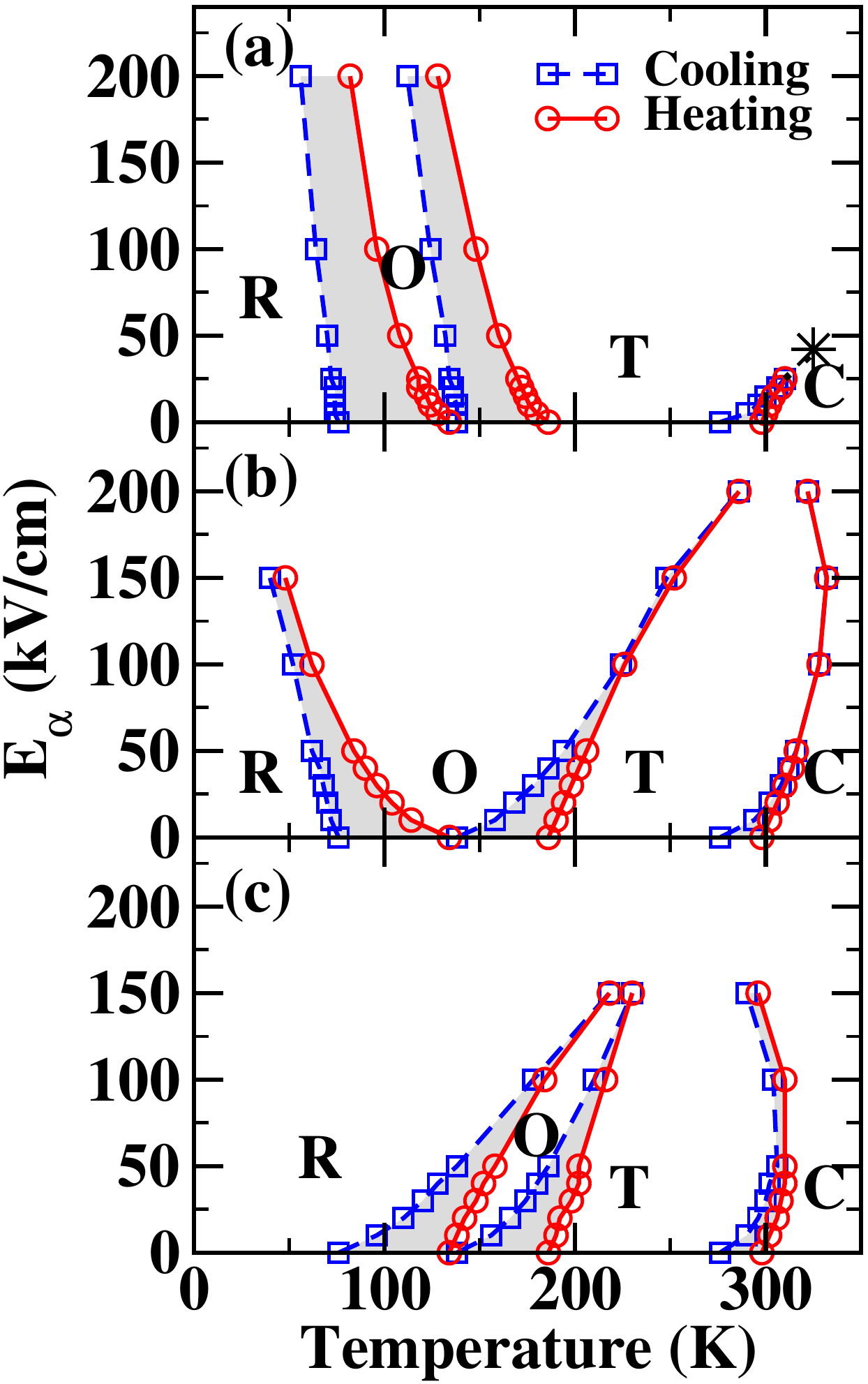}
\caption{Electric field versus temperature ($E$-$T$) phase diagram of
  BTO for fields applied along (a) [001], (b) [011], and (c)
  [111]. Here, $E_{\alpha}$ is the magnitude of each non-zero
  Cartesian component of the applied field. Shaded regions between
  heating and cooling curves correspond to the coexistence regions
  between two phases. The black star in (a) indicates the estimated
  critical point ($E_c, T_c$) for the PE-FE transition with field
  applied along [001].  }
\label{fig:E-Tphasediag}
\end{figure}

The resulting electric field versus temperature ($E$-$T$) phase
diagram is shown in Fig.~\ref{fig:E-Tphasediag}. Each panel
corresponds to the results obtained for a specific direction of the
applied electric field.  The shaded areas between heating and cooling
curves corresponding to the same transition represent coexistence
regions, where the observed phase of the system depends on its
history. For all transitions and field directions, the width of the
thermal hysteresis and thus the coexistence region decreases with
increasing electric field. Note that we expect to obtain a more
pronounced thermal hysteresis compared to what is typically observed
in experiments, because our model system does not include any defects
or inhomogeneities that can act as nucleation centers for the phase
transition.

The overall structure of the $E$-$T$ phase diagram for the three
different field directions agrees well with previous results obtained
from thermodynamic modeling using Landau
theory.~\cite{Bell:2001,Li_Cross_Chen:2005} For each field direction,
the temperature range of the FE phase which has its spontaneous
polarization parallel to the applied field is strongly increasing with
the field strength, and the corresponding transition lines shift to
lower/higher temperatures, accordingly. All other phases will
eventually vanish for sufficiently high field strength. However, the
required fields for the R (R and O) phase(s) and field along [011]
([001]), i.e., the phase(s) on the low temperature side of the phase
favored by the applied field, are much higher than our strongest
applied electric fields. In contrast, the disappearance of the T phase
for strong applied fields along [011] and [111] and of the O phase for
strong applied fields along [111] can already be anticipated from the
$E$-$T$ phase diagram shown in Fig.~\ref{fig:E-Tphasediag}, even
though the corresponding phase regions are not yet completely closed
towards high $E_\alpha$.

For $\vec{E} \parallel [001]$, the first order transition emanating
from the zero-field PE-C to FE-T transition ends in a critical point
$(E_c, T_c)$, see, e.g.,
Ref.~\onlinecite{Novak_Pirc_Kutnjak:2013}. This is because, as already
mentioned, the symmetry of the ``PE'' phase~\footnote{We note that for
  non-zero $E_\alpha$ the distinction between PE and FE phases
  becomes, strictly speaking, meaningless, since there is a nonzero
  polarization at all temperatures. Nevertheless, we use the term
  ``PE'' to indicate the phase emanating from the zero field PE phase
  at high temperatures, which is still separated from the T phase at
  lower temperatures by a first order phase transition, at least for
  not too high $E_\alpha$.}  becomes identical to the one of the T
phase for a field applied along [001]. Simultaneously, the applied
field reduces the jump in $P_z$ associated to the first order phase
transition. Thus, once the temperature dependence of $P_z$ becomes
continuous, no phase transition occurs.

We can obtain a rough estimate for $E_c$ and $T_c$ by linearly
extrapolating the jump in polarization obtained for small electric
fields, i.e., where $\Delta P$ is sizable, towards higher fields, and
then identifying the field and temperature where this extrapolation
becomes zero. In this way, we obtain a critical field around 40\,kV/cm
and a critical temperature around 325\,K, which is indicated by the
black star in Fig.~\ref{fig:E-Tphasediag}(a).  This value obtained for
$E_c$ is larger than what has been measured
experimentally,~\cite{Novak_Pirc_Kutnjak:2013} consistent with the
expected stronger first order character of the transitions in our
simulations compared to experiments (as already mentioned earlier).
Nevertheless, we note that the thermal hysteresis of the corresponding
PE-FE transition in our simulations essentially vanishes already for a
field of around $E_\alpha = 25$\,kV/cm, and no clear jump in $P_z(T)$
is recognizable any more above this field strength (see
Fig.~\ref{fig:P_T-Eapp}a). However, due to the limited numerical
accuracy, finite-size effects, and the discretized $T$-sampling, it is
not straightforward to identify the exact field strength at which
$P(T)$ becomes continuous from our data.

\subsection{Electrocaloric entropy change across the first order phase transitions}
\label{subsec:cc}

Following
Refs.~\onlinecite{Planes/Manosa/Acet:2009,Planes/Castan/Saxena:2014}
we can obtain an expression for the isothermal electrocaloric entropy
change, \mbox{$\Delta S = S(T,\vec{E}) - S(T,\vec{E}=0)$}, under
application of an external electric field, $\vec{E}$, across a first
order phase boundary. For this, we express the field- and
temperature-dependent polarization component along the field direction
in the vicinity of a first order phase transition as follows:
\begin{equation}
\label{eq:pol}
P(T,E) = \tilde{P}(T,E) + \Delta P(E_t(T)) \, \Theta(T-T_t(E)) \ .
\end{equation}
Here, $\tilde{P}(T,E)$ is the part of $P$ that varies smoothly with
$T$ and $E=|\vec{E}|$, $\Delta P(E_t(T))$ is the jump in polarization
at the temperature-dependent transition field $E_t(T)$ (or,
equivalently, at the field-dependent transition temperature $T_t(E)$),
and $\Theta(T-T_t(E))$ is the Heaviside step function, which is 0 for
$T-T_t(E)<0$ and 1 for $T-T_t(E)>0$. Thus, positive (negative) $\Delta
P$ corresponds to the case where the polarization along the field
direction is larger (smaller) above the transition temperature than
below the transition temperature. Furthermore, we assume that $E_t(T)$
and accordingly $T_t(E)$ are uniquely defined in the considered
temperature and field range.

Using the Maxwell relation $(\partial S/\partial E)_T = (\partial
P/\partial T)_E$, where $P$ refers to the polarization component along
the direction of the applied field, and focusing only on the
contribution to $\Delta S$ resulting from the jump in $P(T,E)$, one
obtains:
\begin{equation}
\label{eq:cc}
\Delta S = \Delta P(E_t(T)) \,\left| \frac{dE_t}{dT} \right| \ ,
\end{equation}
which corresponds to the well-known Clausius-Clapeyron
equation.\footnote{Note that the sign convention used here for $\Delta
  P$ is opposite to the one used in
  Ref.~\onlinecite{Planes/Manosa/Acet:2009} for $\Delta M$ and thus
  Eq.~\eqref{eq:cc} has opposite sign compared to Eq.~(11) from
  Ref.~\onlinecite{Planes/Manosa/Acet:2009}.} From Eq.~\eqref{eq:cc}
it follows that the sign of the caloric entropy change when crossing a
first order phase transition is determined by the sign of $\Delta
P(E_t(T))$, i.e. the corresponding jump in the polarization along the
direction of the applied field.

\begin{figure}
\centering 
\includegraphics[width=0.4\textwidth]{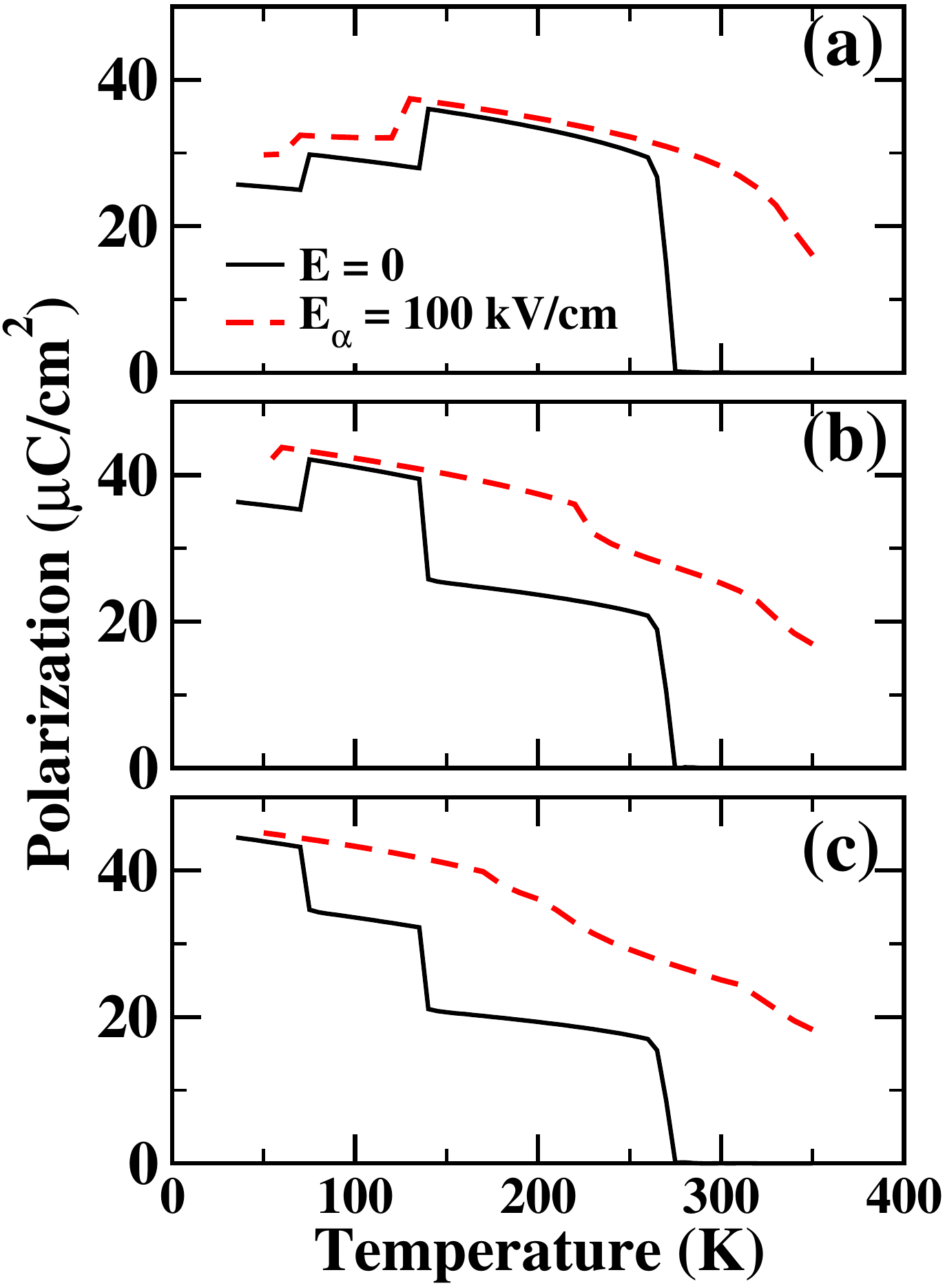}
\caption{Polarization projected along the field direction for
  different directions of the applied field -- (a) [001], (b) [011],
  and (c) [111] -- plotted as a function of temperature for zero field
  (black solid lines) and for $E_\alpha = 100$\,kV/cm (red dashed
  lines). We only show here results obtained during cooling
  simulations. }
\label{fig:P_proj_T3dir}
\end{figure}

Fig.~\ref{fig:P_proj_T3dir} shows the calculated polarization
component along the field direction for different field directions as
function of temperature and for two different field strengths. It can
be seen that both positive and negative $\Delta P$ occurs at the T-O
and O-R transitions, depending on the direction of the applied
field. The polarization jumps are most pronounced for zero field, but
are still qualitatively similar for $E_\alpha = 100$\,kV/cm.
According to Eq.~\eqref{eq:cc}, a negative $\Delta P$ corresponds to
negative $\Delta S$ and thus to a \emph{normal} caloric effect, i.e.,
the entropy is decreased by the applied field. In contrast, positive
$\Delta P$ corresponds to positive $\Delta S$ and thus gives rise to
an \emph{inverse} caloric effect, i.e., the applied field increases
the entropy of the system.

Comparing Fig.~\ref{fig:P_proj_T3dir} with
Fig.~\ref{fig:E-Tphasediag}, one realizes that the cases with positive
$\Delta P$ (T-O and O-R transitions for field along [001] and O-R
transition for field along [011]), and thus with inverse caloric
effect, correspond to the transitions and field directions with
negative $dE_t/dT$, i.e., the cases where the applied field shifts the
transition temperature to lower values.  In these cases, if the
electric field is increased under isothermal conditions, the phase
transition line is crossed coming from the ``low temperature phase''
and entering into the ``high temperature phase'', which is stabilized
at the given temperature due to the applied field. Since the high
temperature phase usually has the higher entropy, this means that
$\Delta S$ is positive, consistent with Eq.~\eqref{eq:cc}. Thus,
Eq.~\eqref{eq:cc} is crucial to understand the sign of the EC response
at the FE-FE transitions, and both positive $\Delta P$ and negative
$dE_t/dT$ indicate an inverse electrocaloric effect at the
corresponding transition for a given field direction. \footnote{Note
  that a negative slope $dE_t/dT$ does not affect the sign of $\Delta
  S$ in Eq.~\eqref{eq:cc} since $\Delta S$ depends only on the
  absolute value of $dE_t/dT$.}  Note that the generalized
Clausius-Clapeyron equation and its implications have already been
discussed in the context of the magnetocaloric effect in shape memory
Heusler alloys.~\cite{Planes/Manosa/Acet:2009}

We can also estimate the EC entropy change $\Delta S$ associated with
the first-order phase transition using our calculated field-dependent
transition temperatures and polarization projections in
Eq.~\eqref{eq:cc}. From the data shown in Fig.~\ref{fig:E-Tphasediag}
we can extract $dE_t/dT$ noting that $T_t(E)$ is approximately linear
at small fields ($\leq 20$\,kV/cm). We therefore obtain
$\text{d}E_t/\text{d}T|_{E=0}$ from a linear fit by using the average
of $T_t(E)$ calculated from heating and cooling simulations. We
estimate the jump in polarization $\Delta P|_{E=0}$ at each transition
by averaging the data shown in Fig.~\ref{fig:P_proj_T3dir} along with
that obtained from the heating simulations.  From Eq.~\eqref{eq:cc} we
then obtain $|\Delta S (E = 0)| = 3.6$\,J/kg/K for the PE-FE
transition, $|\Delta S (E = 0)| = 2.6$\,J/kg/K for the T-O transition,
and $|\Delta S (E = 0)| = 1.4$\,J/kg/K for the O-R transition. We note
that these values are only rough estimates based on our data obtained
for different field directions. Nevertheless, our estimates of $\Delta
S$ compare reasonably well with values of $2.2\pm0.2$\,J/kg/K measured
in Ref.~\onlinecite{Moya_et_al_2013} for the C-T transition, and with
$2.4\pm0.2$\,J/kg/K and $2.2\pm0.2$\,J/kg/K measured in
Ref.~\onlinecite{Stern-Taulats_et_al:2016} for the C-T and T-O
transitions, respectively. A successive reduction of $\Delta S$ for
the \mbox{C-T}, \mbox{T-O}, and \mbox{O-R} transitions has also been
reported in Ref.~\onlinecite{Shirane_et_al:1954}.

\subsection{Direct calculation of EC temperature change}
\label{subsec:deltaT}
\begin{figure}[tb]
\centering \includegraphics[width=0.4\textwidth]{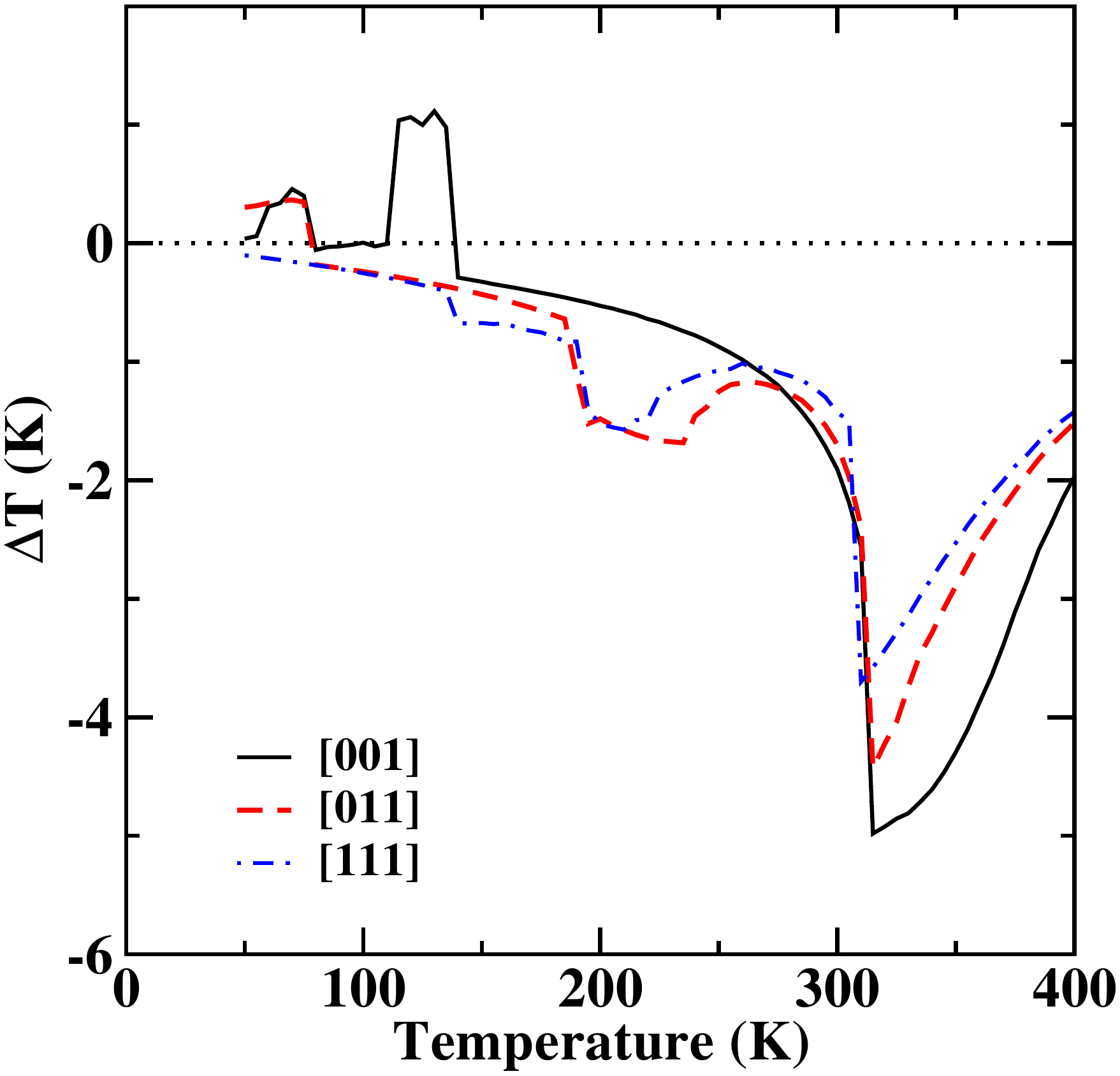}
\caption{EC temperature change under removal of an electric field,
  plotted as function of the initial temperature, for different field
  directions. For each direction the initially applied total field was
  $E = 200$\,kV/cm.}
\label{fig:deltaT-dir}
\end{figure}

Next, we directly calculate the adiabatic temperature change $\Delta
T$ for the three different field directions using the microcanonical
ensemble, as described in Sec.~\ref{sec:method}. In addition to the
contribution resulting from the discontinuous jump ($\Delta P$) at the
phase transition, which was discussed in the previous subsection, the
directly calculated $\Delta T$ also includes the contribution to the
EC response resulting from the smooth variation of $P(T,E)$, i.e,
$\tilde{P}(T,E)$ in Eq.~\eqref{eq:pol}.

The calculated $\Delta T(T_i)$, corresponding to removal of an
electric field at initial temperature $T_i$ is shown in
Fig.~\ref{fig:deltaT-dir} with the same total initial field $E =
|\vec{E}| = 200$\,kV/cm applied along different directions. One can
see that there are sharp features visible in $\Delta T$, corresponding
to all three transitions, regardless of the direction of the applied
field.  The largest response is observed just above the zero-field
PE-FE transition temperature, i.e., slightly above 300\,K, with a
maximum temperature change of around $-5$\,K for $E = 200$\,kV/cm
applied along $[001]$.  This is similar to what has been reported
previously.~\cite{Nishimatsu_Barr_Beckman:2013,Ponomareva/Lisenkov:2012}
In the temperature range of the PE-FE transition and above, the
calculated EC temperature change is qualitatively similar for all
three directions for the applied field, but the magnitude of $\Delta
T$ depends strongly on the field direction and is largest (for the
same total applied field) for $\vec{E} \parallel [001]$.

At lower temperatures, i.e, in the temperature range of the FE-FE
transitions, there are pronounced qualitative differences between the
three different field directions. In each case, $\Delta T$ exhibits
two rectangularly-shaped peaks corresponding to the T-O and O-R
transitions, respectively. However, since the transition temperatures
are generally shifted in different ways for the different field
directions, these peaks in $\Delta T$ appear at different temperatures
in each case. For $\vec{E} \parallel [111]$ the two peaks are
essentially merged into one single feature, since the two FE-FE
transitions are very close in temperature.

Most strikingly, an \emph{inverse} caloric effect, i.e., a positive
$\Delta T$ under field removal, can be observed at the T-O and O-R
transitions for $\vec{E} \parallel [001]$, and at the O-R transition
for $\vec{E} \parallel [011]$. These are exactly those cases where an
inverse isothermal caloric entropy change has been predicted in the
previous subsection based on Eq.~\eqref{eq:cc} and the calculated
jumps in polarization (Fig.~\ref{fig:P_proj_T3dir}). Thus, the
qualitative considerations discussed in the previous subsection, which
were based on Eq.~\eqref{eq:cc}, are consistent with the directly
calculated temperature changes shown in Fig.~\ref{fig:deltaT-dir}.

We note that an earlier study has suggested that the inverse EC effect
occurring close to a FE-FE phase boundary in
Ba$_{0.5}$Sr$_{0.5}$TiO$_3$ originates from noncollinearity between
polarization and electric field.~\cite{Ponomareva/Lisenkov:2012} Our
results offer further clarification. In particular, one can see that
$\Delta T$ remains negative, i.e., ``normal'', over the whole
temperature range for $\vec{E} \parallel [111]$, in spite of the
misalignment between the applied field and the polarization in both
the T and O phases in this case. Thus, even if the applied field is
not parallel to the polarization in any of the FE phases on either
side of the transition, a normal EC effect can occur. The discussion
in the previous section based on the Clausius Clapeyron equation
offers a simple explanation of this, and shows that the crucial point
is that the applied field induces a transition from a ``low entropy''
to a ``high entropy'' FE phase. While the noncollinearity between the
field and the polarization direction in at least one of the FE phases
is necessary to induce such a transition, the presence of the
noncollinearity by itself is not sufficient to cause an inverse
caloric effect.

\begin{figure}[tb]
\centering \includegraphics[width=0.4\textwidth]{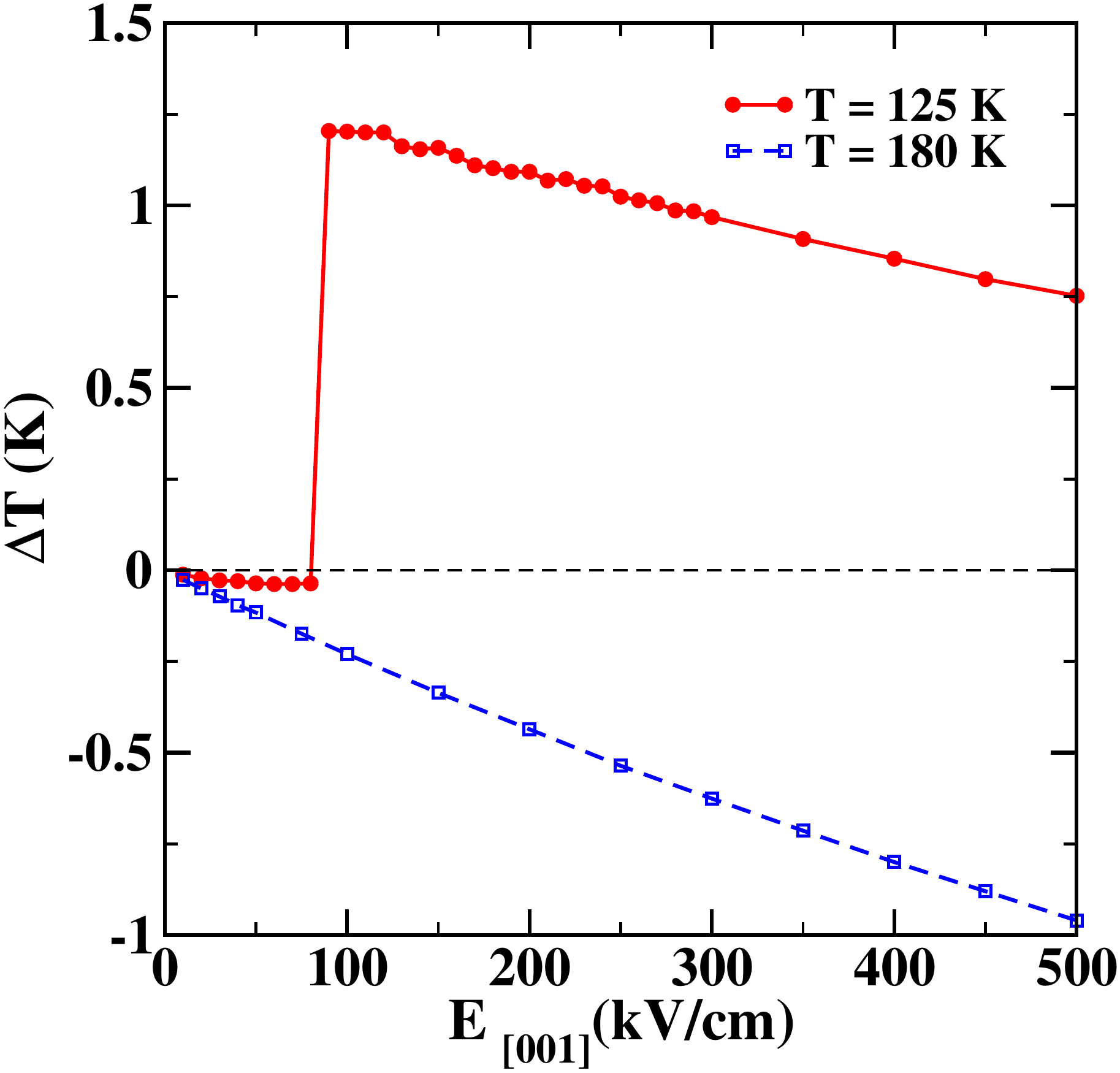}
\caption{Adiabatic EC temperature change, $\Delta T$, as function of
  the initially applied field magnitude for $\vec{E} \parallel
  [001]$. Each data point is calculated separately as the field is
  switched-off from the specified initial temperature, as shown
  schematically in Fig.~\ref{fig:delT-scheme}.}
\label{fig:deltaT-mag}
\end{figure}

In order to further analyze the occurrence of the inverse caloric
effect, we now focus on the temperature region around the T-O
transition for an applied field along the [001] direction, and we
calculate the adiabatic EC temperature change at a fixed initial
temperature but for different strengths of the applied electric
field. In Fig.~\ref{fig:deltaT-mag} we show the corresponding results
for two different initial temperatures, $T_i=180$\,K and $T_i=125$\,K.

As seen from Fig.~\ref{fig:E-Tphasediag}(a), at $T_i=180$\,K, the T
phase is stable over the whole range of applied field strengths, even
for zero applied field. Consequently, the system is initially in the T
phase and remains so under field removal. Fig.~\ref{fig:deltaT-mag}
shows that this results in a normal, i.e., negative, EC temperature
change, the magnitude of which increases monotonously with increasing
field strength.

In contrast, for an initial temperature of 125\,K, the T phase is
unstable for applied fields below $\sim$90\,kV/cm, i.e., below the
blue T-O transition line in Fig.~\ref{fig:E-Tphasediag}(a). Thus, if
for large applied fields the system is initially in the T phase it
transforms into the O phase under field removal. As seen from
Fig.~\ref{fig:deltaT-mag}, this results in a distinctly non-monotonous
dependence of $\Delta T$ on the applied field strength.  For small
applied fields (below $\sim$90\,kV/cm), the system is in the O phase
both with and without applied field and exhibits a (rather small)
normal EC effect (negative $\Delta T$) that increases slightly with
the field strength. However, once the applied field becomes larger
than $\sim$90-100\,kV/cm, the system is initially in the T phase but
then transforms into the O phase under field removal. This phase
transition results in a strong inverse EC effect with a (positive)
adiabatic temperature change of around 1.2\,K, consistent with the
negative (i.e, inverse) EC entropy change obtained from
Eq.~\eqref{eq:cc} for the T-O transition and field along [001].
If the magnitude of the applied field is further increased, then
$\Delta T$ decreases again, indicating a normal EC effect within the T
phase. Thus, the inverse effect is indeed closely related to the T-O
phase transition, as outlined in the previous subsection, and occurs
exactly at the specific temperature-dependent transition field
$E_t(T)$ where the transition line is crossed.

\begin{figure}
\centering \includegraphics[width=0.4\textwidth]{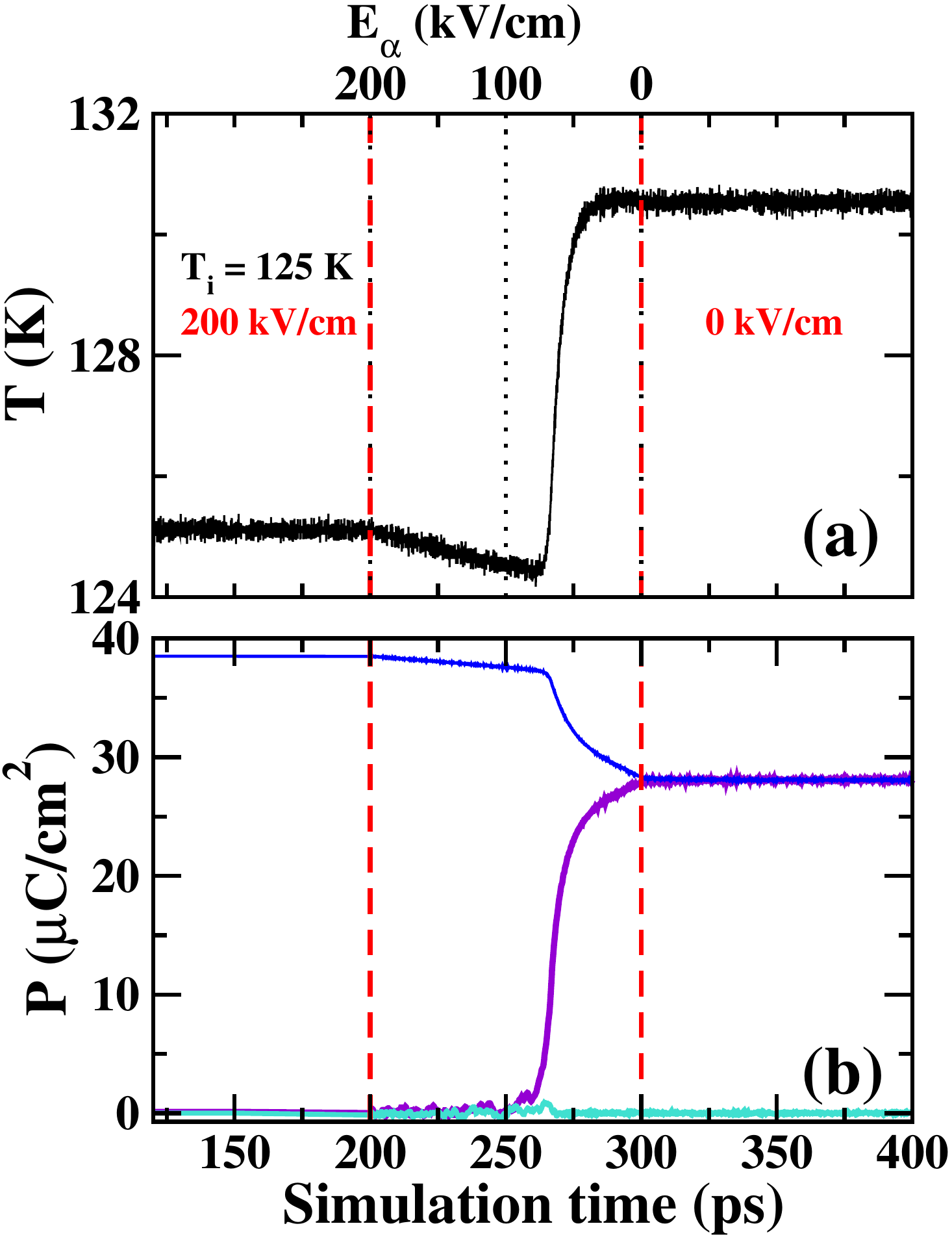}
\caption{Time evolution of the system temperature (a) and of the
  Cartesian components of the polarization (b) during the course of a
  simulation with an initial temperature of 125\,K and an initial
  applied field of 200\,kV/cm along [001]. Between $t=200$\,ps and
  $t=300$\,ps the applied field is ramped down to zero, as indicated
  on the top side of the graph.}
\label{fig:T125-time}
\end{figure}

Fig.~\ref{fig:deltaT-mag} contains the \emph{total} $\Delta T$ under
removal of a given initial field. To complement this, we can also plot
the actual time evolution of the temperature (defined through the
average kinetic energy) while we remove the field during the course of
the simulation. This is shown, together with the evolution of the
polarization components, in Fig.~\ref{fig:T125-time}. The system is
thermalized at $T_i = 125$\,K and an initially applied field of
200\,kV/cm along [001].  It can be seen, that at the start of the
simulation the system is in the T phase, with only one non-zero
component of $P$. When the field is ramped down (between $t=200$\,ps
and $t=300$\,ps), this component starts to decrease with the field,
and simultaneously we observe a weak decrease in temperature, i.e., a
normal EC effect.  This confirms that the decrease of $\Delta T$
observed in Fig.~\ref{fig:deltaT-mag} for $T_i=125$\,K and applied
fields above 90\,kV/cm is indeed related to the (relatively weak)
normal EC effect within the T phase, that occurs while the initially
applied field is ramped down to a value of around 90\,kV/cm (see the
evolution of $T$ between $t=200$\,ps and $t=250$\,ps in
Fig.~\ref{fig:T125-time}).

However, once the field has decreased below 90\,kV/cm, the T-O phase
transition sets in, and the system starts to transform into the O
phase, characterized by two equal non-zero polarization
components. This transformation is accompanied by a sharp increase in
temperature, corresponding to an inverse caloric effect.  At the end
of the field ramping phase ($t=300$\,ps), the system has completely
evolved into the O-phase and the overall change in temperature
corresponds to an inverse EC effect.

As seen from Fig.~\ref{fig:T125-time}, during the course of the phase
transition, the system temperature in our simulation changes by about
6\,K, which, after rescaling to the correct number of degrees of
freedom, corresponds to $\Delta T \approx 1.2$\,K, consistent with the
temperature change obtained from Fig.~\ref{fig:deltaT-mag}.  It is
important to note that, since this temperature change occurs exactly
when the system is crossing the transition line, its magnitude is
essentially independent of the strength of the initially applied
electric field. The only requirement is that the applied field is
strong enough to cross the first order T-O transition line for a given
initial temperature. If the transition occurs, it results in an
inverse caloric temperature change of around 1\,K.

\begin{figure}[tb]
\centering
\includegraphics[width=0.35\textwidth]{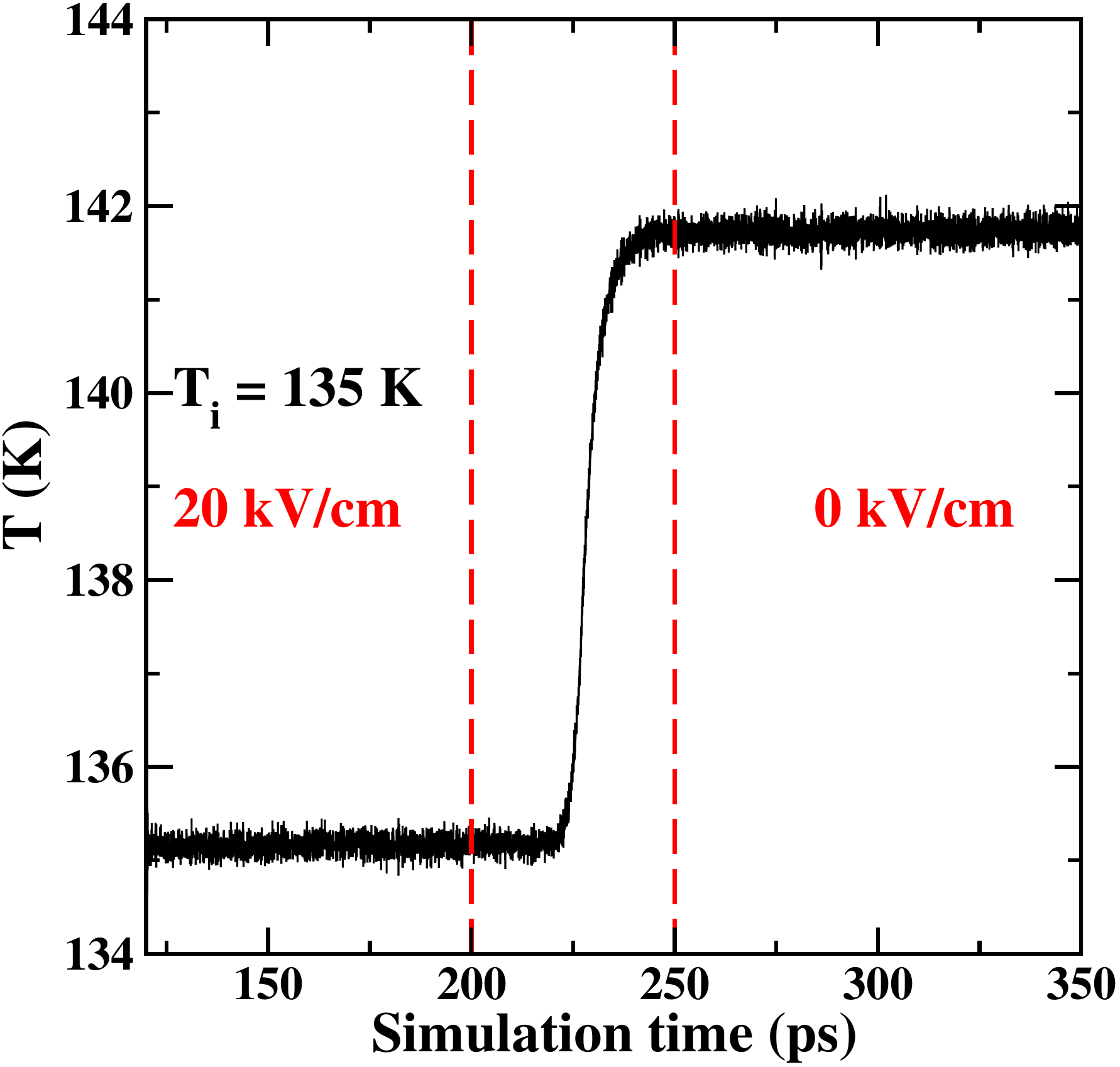}
\caption{Time evolution of the system temperature during a simulation
  for $T_i = 135$\,K, when a field of 20\,kV/cm along $[001]$ is
  switched off. The time period between the two red dashed lines
  indicates the time during which the field is slowly ramped down with
  a rate of $-0.0004$\,kV/cm/fs.}
\label{fig:T-O-latheat}
\end{figure}

For $T_i=125$\,K, an applied field of at least 100\,kV/cm is required
to cross the T-O transition. However, the transition field is strongly
reduced for slightly higher temperatures. For example, at $T_i =
135$\,K the T-O transition can already occur for an applied field of
20\,kV/cm, and indeed we obtain essentially the same inverse
temperature change as for $T_i=125$\,K when the system undergoes the
transition (see Fig.~\ref{fig:T-O-latheat}).

We note that in the present case, if the system is initially within
the coexistence region, we initialize the system such that the
application or removal of the field will result in a crossing of the
transition line we want to study. However, without initial poling of
the system or upon subsequent field cycling, the system may get stuck
in one of the FE phases, and thus will not undergo a phase transition
in subsequent field cycles. Furthermore, the system can also form
complex multi-domain states,~\cite{Limboeck:Soergel:2014} which can,
potentially, reduce the overall EC response. In practice, the limiting
factor for obtaining the large EC effect related to the T-O transition
entropy, is that the applied field has to be large enough to overcome
the hysteretic behavior of the system, in order to reversibly drive
the system back and forth between the two different phases.

\subsection{Direct measurements of the EC temperature change in BTO single crystals}
\label{subsec:exptdeltaT}

In order to verify the theoretical considerations and simulation
results presented in the preceding subsections, we perform direct
measurements of the EC temperature change in single crystals of BTO
with different orientation of the crystallographic axes relative to
the applied electric field (see Sec.~\ref{sec:exp} for details on the
experimental procedure).  The EC measurements are performed in the
vicinity of both the cubic-tetragonal (C-T) and the
tetragonal-orthorhombic (T-O) phase transitions.

\begin{figure}[tb]
\centering
\includegraphics[width=0.4\textwidth]{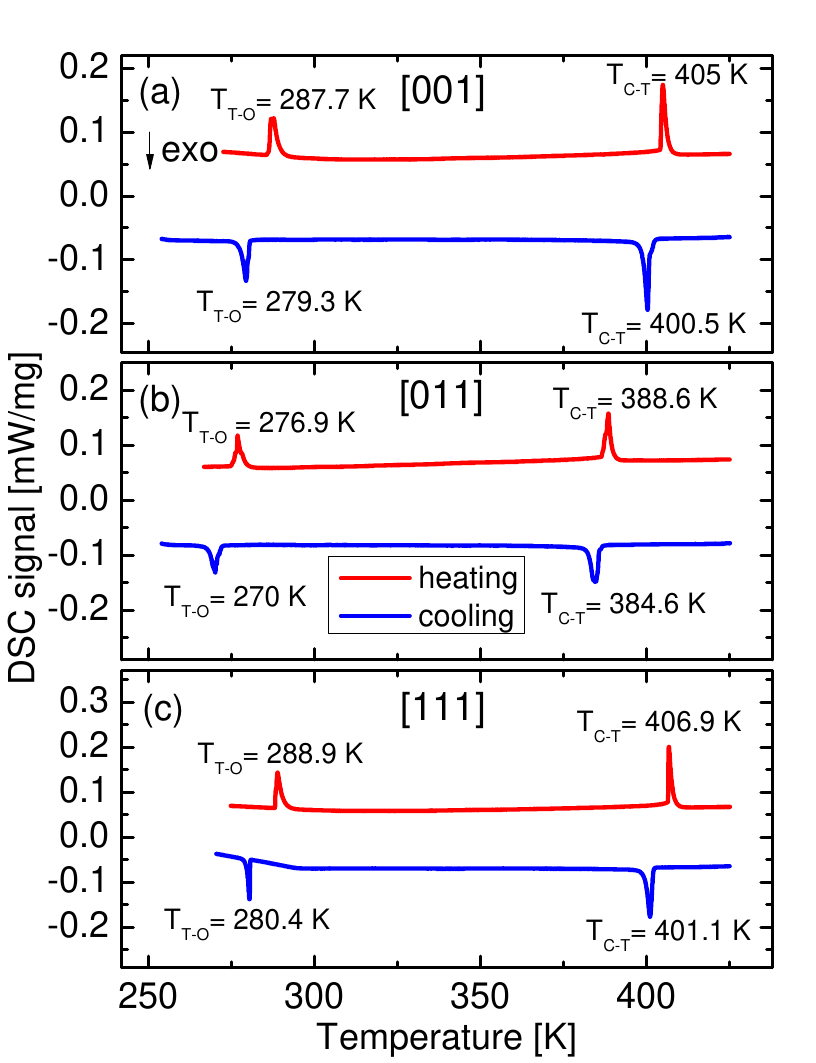}
\caption{Results of differential scanning calorimetry (DSC)
  measurements to obtain the phase transition temperatures for (a) BTO
  [001], (b) BTO [011], and (c) BTO [111] on cooling (blue) and on
  heating (red).}
\label{fig:DSC}
\end{figure}

In Fig.~\ref{fig:DSC}, we first show the DSC signals measured for all
BTO samples on both heating and cooling, from which the transition
temperatures for the C-T and T-O transitions can be obtained. Note
that in all cases, there is a thermal hysteresis of a few Kelvin
between heating and cooling at both transitions. Furthermore, there is
a slight shift in the transition temperatures for different BTO
samples.  We attribute the variation of the transition temperatures to
possible variations of impurity concentration in the crystals under
study.

\begin{figure}[tb]
\centering
\includegraphics[width=0.4\textwidth]{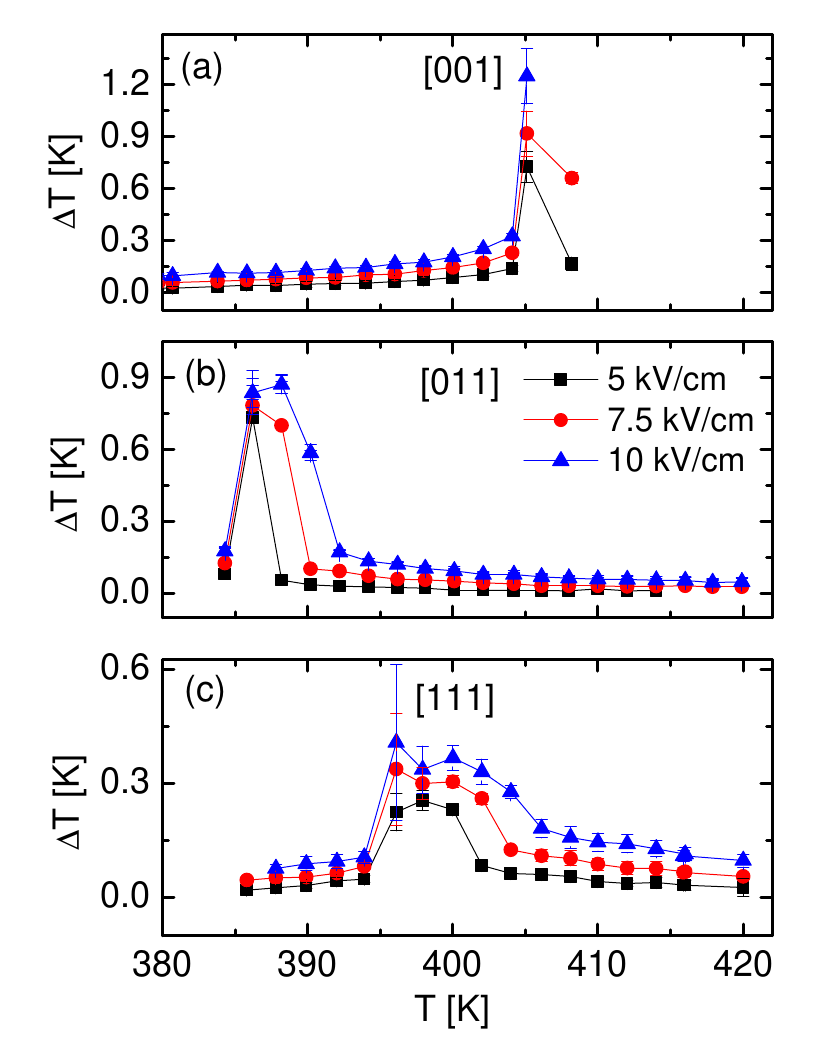}
\caption{Directly measured electrocaloric temperature change $\Delta
  T$ for three different BTO single crystal orientations (a) BTO
  [001], (b) BTO [011] and (c) BTO [111] for different applied
  electric fields in the vicinity of the cubic-tetragonal (PE-FE)
  phase transition. The error bar at each point is equal to the
  calculated standard deviation.}
\label{fig:expt_C-T}
\end{figure}

Fig.~\ref{fig:expt_C-T} shows the measured EC temperature changes for
the three BTO single crystals with different orientation near the C-T
(PE-FE) phase transition. One can see that for all three orientations
we obtain a positive EC effect, and the position of maximum $\Delta T$
correlates well with the transition temperatures estimated from the
DSC measurements (see Fig.~\ref{fig:DSC}).~\footnote{We note that
  there is often a discrepancy between transition temperatures
  estimated using different methods. This can be related to
  differences in experimental setups such as the rate of temperature
  change, the position of the temperature sensor relative to the
  sample, etc.}  Away from the phase transition region, $\Delta T$ is
significantly reduced.  The largest EC temperature change is obtained
for the BTO single crystal oriented along the [001] direction, with
$\Delta T_{\text{max}} = 1.25$\,K at 405\,K for an applied field of
10\,kV/cm. The magnitude of $\Delta T_{\text{max}}$ decreases for the
[011]-oriented crystal (0.87\,K at 388\,K) and then further for the
[111]-oriented sample (0.4\,K around 400\,K) under the same applied
field. For the [001] oriented sample, measurements at temperatures
$\ge 410$\,K were not possible due to large leakage currents.

This trend agrees well with what is observed in our MD simulations,
i.e., for the same total applied field, the EC response is largest
along [001] and smallest along [111].  Further, we note that the fact
that the low temperature side of the measured EC peak is essentially
field independent, whereas the high temperature side is shifted
towards higher temperatures with increasing field strength, also
agrees well with our simulations.

\begin{figure}[tb]
\centering
\includegraphics[width=0.4\textwidth]{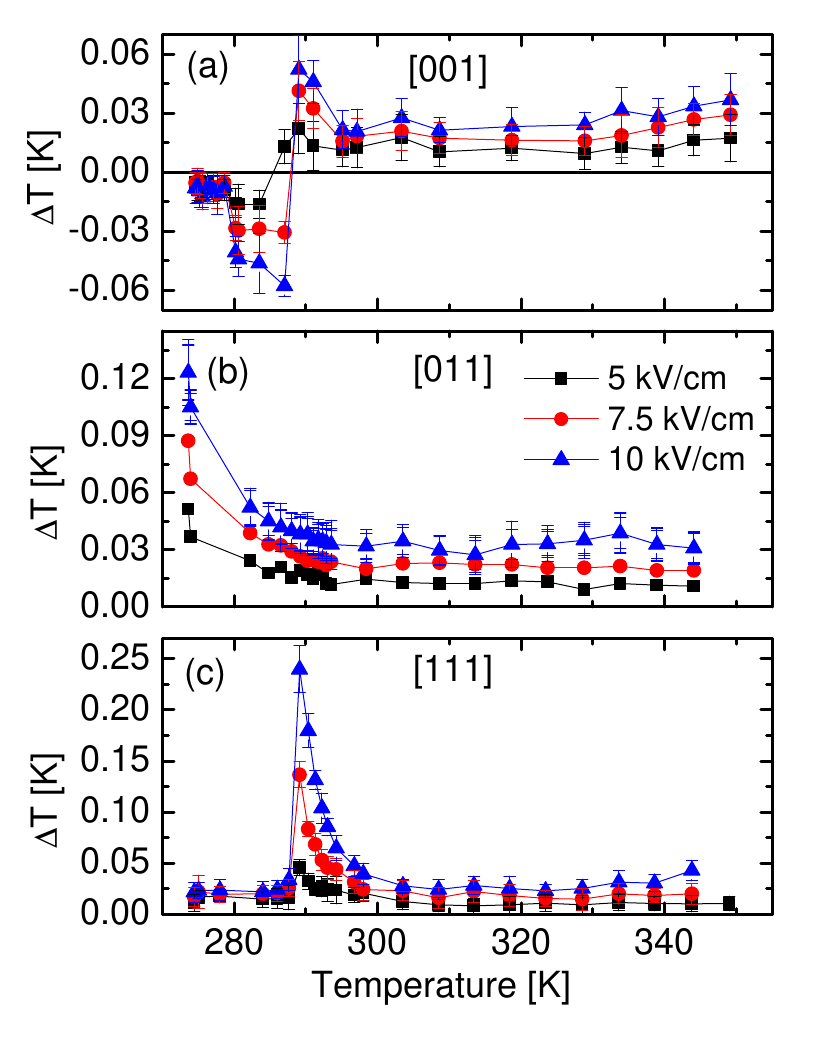}
\caption{Directly measured electrocaloric temperature change $\Delta
  T$ for three different BTO single crystal orientations (a) BTO
  [001], (b) BTO [011] and (c) BTO [111] for different applied
  electric fields in the vicinity of the tetragonal-orthorhombic
  (FE-FE) phase transition. }
\label{fig:expt_T-O}
\end{figure}

Fig.~\ref{fig:expt_T-O} depicts the EC temperature change of the three
samples close to the T-O (FE-FE) phase transition. Again, the
temperatures where the maximal $\Delta T$ values can be observed are
consistent with the O-T transition temperatures obtained from the DSC
measurements. A normal (positive) EC effect is observed for both the
[011] and [111]-oriented crystals. In contrast, an inverse (negative)
EC effect occurs for the [001]-oriented crystal, which switches to a
positive $\Delta T$ for increasing temperature. The inverse EC effect
in the [001] sample reaches a magnitude of $\Delta T_{\text{max}} =
-0.06$\,K for an applied electric field of $10$\,kV/cm in the vicinity
of the transition, whereas at lower temperatures it is very small and
field-independent.  For the [011] and [111]-oriented crystals the EC
effect reaches maximum values of $\Delta T_{\text{max}} = 0.12$\,K and
0.24\,K, respectively. We note that our current setup does not allow
to cool below 273\,K, and therefore we could only measure the high
temperature side of the $\Delta T$ peak for the [011] sample.

The inverse response measured for the [001] sample (and the positive
response for the other two orientations) is in excellent qualitative
agreement with our theoretical predictions for the EC response at the
O-T transition in BTO. However, quantitatively, the measured
temperature changes are much smaller than what is predicted by our
simulations. Furthermore, in the simulations, $|\Delta T_\text{max}|$
was rather similar for all three field directions (since it is closely
related to the T-O transition entropy), whereas in the experiments
$|\Delta T_\text{max}|$ varies between different field
directions. These quantitative differences are most likely due to the
smaller fields applied in the experiments, which are perhaps not
sufficient to completely drive the system through the O-T
transition. Furthermore, the presence of multiple domains in the
experimentally studied samples, the influence of defects, impurities,
thermal losses, etc., can in principle account for the quantitative
differences between the experimental results and the simulations,
which are performed for an ideal model system.
 
We point out that Bai, \textit{et al.}~\cite{Bai_et_al:2012} reported
$\Delta T$ = 1.4\,K under 10\,kV/cm for BTO single crystals at the T-O
transition, which agrees well with the magnitude of $\Delta T$
predicted from our simulations.  However, this large response was
obtained only during the first field switching, and was reduced by up 
to two orders of magnitude on further cycling. This is in accordance 
with the argument, that for small fields the system may get stuck within 
the coexistence region or a multi-domain configuration, reducing the 
overall response of the system.  In our measurements, we do not observe 
a significant difference between first and subsequent field cycles. 
This could be attributed to different measurement protocols and sample 
histories compared to Ref.~\onlinecite{Bai_et_al:2012}. However, further 
experiments are required to fully resolve these issues.

\section{Summary and Conclusions}
\label{sec:summary}

In summary, we have used molecular dynamics simulations for a first
principles-based effective Hamiltonian to study the EC effect in
BaTiO$_3$ at all three ferroelectric transitions.  Thereby, we have
focused in particular on how the EC effect depends on the direction of
the applied electric field relative to the crystallographic axes of
BaTiO$_3$, and we have analyzed the occurrence of an inverse EC effect
at the FE-FE transitions for certain field directions.  We have also
performed direct measurements of the EC temperature change for BTO
single crystals with electric fields applied along different
crystallographic directions, in order to verify our theoretical
predictions.

Our results show that there is a pronounced anisotropy of the EC
effect in BaTiO$_3$. For temperatures above the zero-field PE-FE
transition temperature, i.e. where the EC effect is largest, the
temperature dependence is qualitatively similar for all field
directions. Nevertheless, its magnitude depends strongly on the field
direction and is largest for fields applied along [001] (and
crystallographically equivalent directions). This is also confirmed by
our direct EC measurements on BTO single crystals.

In the temperature regions around the FE-FE transitions, the EC
response exhibits pronounced qualitative differences for the different
field directions.  Most strikingly, an inverse EC effect is observed
for certain orientations of the applied field.  Using the generalized
Clausius-Clapeyron equation \eqref{eq:cc}, we have shown that such an
inverse EC effect occurs if the electric field induces a transition
from a ``low entropy'' towards a ``high entropy'' FE phase, which can
happen if the polarization in the high entropy phase is oriented more
favorably with respect to the applied field than in the low entropy
phase. This is indicated by a negative slope $dE_t/dT$ of the
corresponding phase transition line in the $E$-$T$ phase diagram, and
leads to a positive jump of the polarization component along the field
direction at the phase transition.

By monitoring the time evolution of polarization and temperature
throughout the simulation, and by calculating $\Delta T$ for different
applied field strengths at selected temperatures, we have shown that in
the simulations the inverse effect occurs exactly when the system
undergoes the phase transformation and is thus intimately related to
the entropy change across the phase transition. Consequently, the
corresponding EC effect is only indirectly affected by the strength of
the applied field, as long as it is sufficient to drive the system
across the transition.

These conclusions are corroborated by direct measurements of the EC
temperature change of BTO single crystals around the T-O transition,
where we indeed observe an inverse EC response for fields applied
along the [001] direction, whereas a normal EC effect is measured for
the other two field directions ([110] and [111]). Thus, there is an
overall very good qualitative correspondence between our theoretical
and experimental results.

Finally, we point out the anisotropy of the EC effect is not only
relevant for single crystals, but has also very important implications
for measurements on polycrystalline samples. In a polycrystalline
sample, different grains will have different orientations of their
crystal axes relative to the applied electric field, and thus only an
average over all possible orientations is measured. Furthermore, if
certain orientations can even lead to an inverse EC effect, this will
lead to partial cancellation and a weakening of the overall EC
response. On the other hand, if the EC response is particularly strong
for certain orientations, this can provide a very efficient route for
optimization of the EC response by using textured polycrystalline
samples, i.e. samples where certain grain orientations are
preferred.~\cite{LeGoupil_et_al:2014}

\section{Acknowledgments}
This work was supported by the Swiss National Science Foundation and
the German Science Foundation (Deutsche Forschungsgemeinschaft, DFG)
under the priority program SPP 1599 (``Ferroic Cooling''), project
codes 200021E-162297, GR 4792/1-2, and Lu729/15. We also thank 
A.~Planes for useful discussions. 

\bibliography{references}

\end{document}